\documentclass[slac_one]{revtex4}
\usepackage{graphicx}
\usepackage{fancyhdr}
\newcommand{\lsim}{\lesssim}

\def\ie{{\it i.e.}}

\def\mpl{\ifmmode M_{pl}\else $M_{pl}$\fi}
\def\mpl{\ifmmode \overline M_{Pl}\else $\bar M_{Pl}$\fi}
\def\to{\rightarrow}
\def\lsim{\mathrel{\mathpalette\atversim<}}

\def\grtsim{\,\,\rlap{\raise 3pt\hbox{$>$}}{\lower 3pt\hbox{$\sim$}}\,\,}
\def\lsim{\,\,\rlap{\raise 3pt\hbox{$<$}}{\lower 3pt\hbox{$\sim$}}\,\,}

\begin{document}

\hfill$\vcenter{
\hbox{\bf MADPH-05-1441}
\hbox{\bf SLAC-PUB-11472}}
$

\title{{\small{2005 ALCPG \& ILC Workshops - Snowmass,
U.S.A.}}\\ 
\vspace{12pt}
Probing the Universal Randall-Sundrum Model at the 
ILC} 

%
%
\author{H. Davoudiasl}
\affiliation{Dept. of Physics, University of Wisconsin, Madison, WI, 53706,USA}
\author{B. Lillie}
\affiliation{SLAC, Stanford, CA 94025, USA}

\author{T.~G. Rizzo}
\affiliation{SLAC, Stanford, CA 94025, USA}

\begin{abstract}
The Randall-Sundrum model with all Standard Model (SM) fields in the 
bulk, including the Higgs, can be probed by precision measurements 
at the ILC. In particular, the couplings of the Higgs to the gauge bosons 
of the SM can be determined with high accuracy at the ILC. Here we examine the 
deviations in these couplings from their SM values within the framework 
of the Universal Randall-Sundrum Model (URSM) as well as the corresponding 
couplings of the first Higgs Kaluza-Klein excitation. 
\end{abstract}

\maketitle

\thispagestyle{fancy}

The Randall-Sundrum (RS) model provides an interesting explanation of the 
hierarchy problem{\cite{Randall:1999ee}}.  In the original RS model, the
only 5-d field is the graviton. Since that time numerous works have 
extended the RS setup to include bulk fermions and gauge 
fields{\cite{Davoudiasl:2005uu}}. However, even though the SM can be promoted 
to a 5-d theory, the fundamental Higgs field responsible for electroweak 
symmetry breaking (EWSB) has been kept on the TeV-brane. The reason for 
this has been to avoid problems associated with extreme fine-tuning and 
conflict with known experimental data. There have also been attempts to 
build RS models {\it without} fundamental Higgs bosons where the role of the 
Higgs doublet as a source of Goldstone bosons is played by the fifth components 
of the gauge fields themselves. 
 
We have recently shown{\cite{Davoudiasl:2005uu}} that by appropriate
choices of the Higgs sector parameters in the bulk and on both the TeV and 
Planck branes, one can generate a single tachyonic Kaluza-Klein (KK) mode of 
the Higgs field in the low energy 4-d theory.  This tachyonic mode can be 
identified as the SM Higgs field. Given a quartic bulk term for the Higgs, 
the tachyonic mode will lead to the usual 4-d Higgs mechanism and endow the 
electroweak gauge bosons with mass. 
A particularly interesting realization of this ``Off-the-Wall'' Higgs EWSB 
employs the gravitational sector of the RS model to provide the necessary bulk
and brane mass scales.  These scales are then related to the 
5-d curvature of the RS geometry.  Consequently, new relations 
and constraints among the parameters of the Higgs and gravitational sectors 
are obtained.  In this gravity-induced EWSB
scenario, the higher curvature Gauss-Bonnet terms play an important role.  
Typical generic signatures of our scenario are  
the emergence of a tower of Higgs KK modes as well as a modification of the 
$WWH$ coupling which can be directly probed through precision measurements at 
the ILC. 
 
In order to understand the basic model let us focus on the free Higgs 
case (\ie, ignoring gauge and self-couplings as well as other
particles such as the Goldstone bosons) which involves only the various mass terms. 
Studying this truncated action for the free Higgs field allows us to perform the KK 
decomposition $\phi \to \sum_n \phi_n(x) \chi_n(y)=\phi_T\chi_T+...$ where $\chi_T$ 
is the wavefunction for the tachyon and represents the shape of the 
Higgs vev profile in the bulk. We have
\begin{equation}
S_{trunc}=\int d^5x ~\sqrt {-g}\Bigg[(\partial^A\phi)^\dagger (\partial_A
\phi)-m^2\phi^\dagger \phi+{1\over {k}}\phi^\dagger \phi~[\mu_P^2\delta(y)-
\mu_H^2\delta(y-\pi r_c)]\Bigg]\,.
\end{equation}
To scale out dimensional factors 
we define $m^2=20k^2\xi$ and $\mu_{P,H}^2=16k^2\xi\beta_{P,H}$ since $k$ is the
canonical scale for RS masses. Recall that the RS metric is given by 
$ds^2=e^{-2k|y|} \eta_{\mu\nu}dx^\mu dx^\nu-dy^2$ where we use 
$-\pi r_c\leq y\leq \pi r_c$ as the 5-d 
coordinate and $k$ is the curvature parameter whose value is comparable to that of 
the fundamental scale $M$. Note that the parameters $\xi, \beta_{P,H}$ are 
dimensionless and are expected to be $O(1)$ but may, in principle, be of arbitrary 
sign. What we looking for in the $(\xi, \beta_{P,H})$ parameter space is to obtain
EWSB in a manner consistent with the SM; our basic criterion is to find
regions where there exists one, and only one, TeV scale
tachyonic mode that we can identify with the SM Higgs. The remaining Higgs KK
tower fields must also be normal, \ie, non-tachyonic. To this end we solve the equation 
of motion for the Higgs KK wavefunctions:
\begin{equation}
\partial_y\Big(e^{-4k|y|}\partial_y \chi_n\Big)-m^2e^{-4k|y|}\chi_n+
{1\over{k}}e^{-4\sigma}[\mu_P^2\delta(y)-\mu_H^2\delta(y-\pi r_c)]\chi_n
+m_n^2e^{-2k|y|}\chi_n=0\,,
\end{equation}
and apply the appropriate boundary conditions 
on both the TeV and Planck branes{{\cite{Davoudiasl:2005uu}}; one can show that 
$\beta_P$ 
plays no essential role in the solutions of interest to us. We find that if the 
combination $\xi \beta_H >0$ then either no tachyon exists or that the resulting 
Higgs vev is Planck scale. Clearly, we must instead choose the parameters such 
that $\xi\beta_H <0$. In this case their are two remaining regions:  
(I)$\xi>0, ~\beta_H<0$ or (II)$\xi<0, ~\beta_H>0$. Furthermore, we find that within 
both of these regions further constraints on the parameters must apply in order that a 
tachyon root exist: $\xi \geq(\leq)\xi_{1,2}$ in region I(II). $\xi_{1,2}$ are
given by the expressions $\xi_1=-{1\over {4\beta_H}}$ and $\xi_2={{5-8\beta_H}\over 
{16\beta_H^2}}$. We note that region II is favored by the ``gravity-induced'' 
EWSB mechanism since in that region it is easy to avoid ghosts in the radion and 
graviton sectors. Clearly the detailed nature of the solutions for the bulk Higgs 
profile $\chi_T$ will depend on the particular values of the parameters 
$\xi,~\beta_H$. One finds that the Higgs profile is highly peaked near the 
TeV brane, $\chi_T \sim e^{(2+\nu)k|y|}$ where $\nu^2=4+20\xi$. The masses of the 
Higgs boson and its KK excitations are found to be given by the roots, $x_n$, of 
a transcendental equation involving Bessel functions of order $\nu$ and depend 
on the values of both $\xi$ and $\beta_H$.

Let us now turn to the gauge boson couplings of the Higgs in URSM. 
In the SM the $W$ boson gets its mass directly from the vev of the Higgs boson. In 
the case of the URSM, the $W$ has an additional, geometric, mass source arising 
solely from back reaction, \ie , wavefunction curvature. This implies that the 
$HWW$ (and, hence, $HZZ$) coupling 
will be somewhat smaller than that obtained in the SM. 
The relevant coupling can be expressed as an integral over $y$:
\begin{equation}
S_{eff}= \int d^4x~HW_\mu^\dagger W^\mu ~{1\over {4\pi r_c}}g^2v \int~dy~
e^{-2k|y|}\chi_T^2 f_0^2\,,
\end{equation}
where $v$ is the Higgs vev, $g$ 
is the usual 4-d coupling assuming that the light SM fermions 
are localized near the Planck brane, and $f_0(y)$ is the $W=W_0$ boson 
KK wavefunction. (It is important to note at this point that $v$ need not be 
$v_{SM} \simeq 246$ GeV.) 
These are 
obtained by KK expanding the 5-d $W$ field $W=\sum_n W_n(x)f_n(y)$ 
and solving the mode equation 
\begin{equation}
\partial_y\Big(e^{-2k|y|}\partial_y f_n\Big)-{1\over {4}} g_5^2v^2\chi_T^2
e^{-2k|y|}f_n+m_n^2f_n=0\,,
\end{equation}
where $g=g_5/{\sqrt {2\pi r_c}}$ is the 
5-d $SU(2)_L$ gauge coupling, with suitable boundary conditions. In 
this language the $W$ boson mass is given by  
\begin{equation}
M_W^2={1\over {4}} 2\pi r_c g^2v^2\int ~dy ~e^{-2k|y|}f_0^2\chi_T^2 +
\int ~dy~e^{-2k|y|}(\partial_y f_0)^2\,. 
\end{equation}
\begin{figure*}[t]
\centerline{
\includegraphics[width=5.8cm,angle=-90]{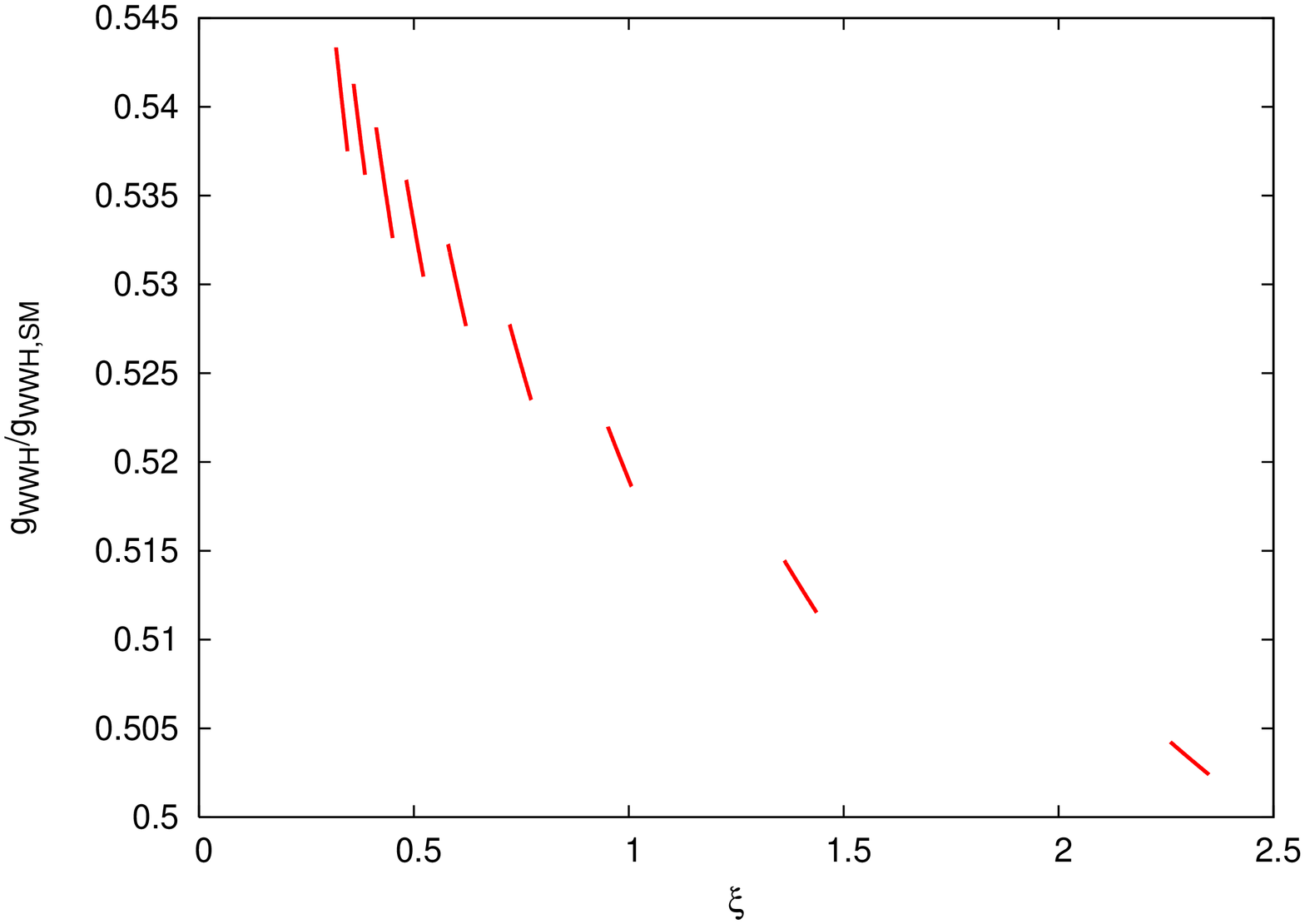}
\hspace*{5mm}
\includegraphics[width=5.8cm,angle=-90]{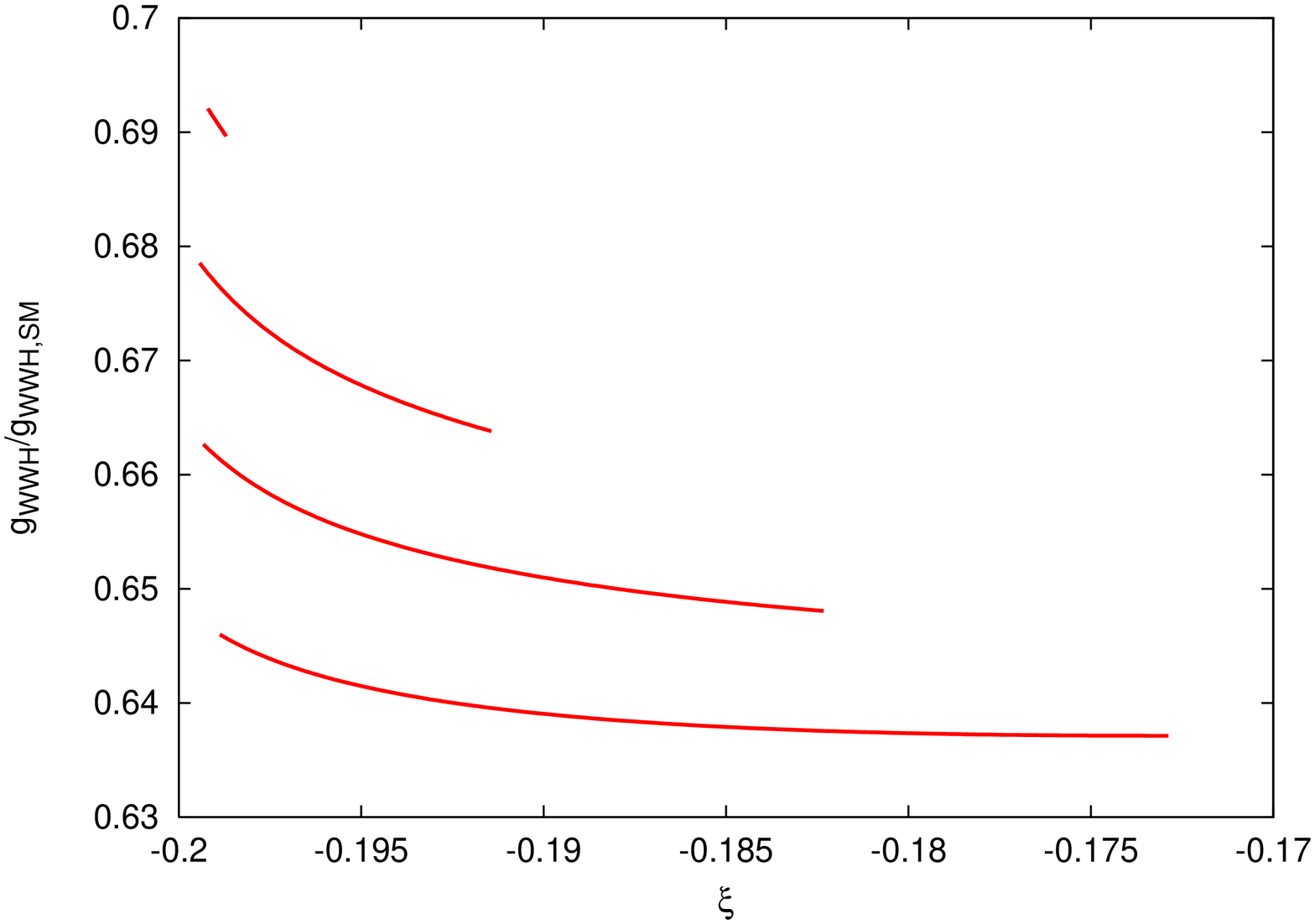}}
\vspace*{0.1cm}
\caption{Values of the coupling ratio $g_{WWH}/g^{SM}_{WWH}$ in regions I (left) 
and 
II (right) as functions of $\xi$ for various $\beta_H$. In region I, from top 
from top to bottom, the curves correspond to $\beta_H=1.4,1.6,1.8,2.0$. A cut on 
the Higgs boson mass as described in the text has been imposed.}
\label{fig1}
\end{figure*}

How do the $WWH$ couplings in the URSM compare to their SM values, \ie, what 
are ratios  
$g_{WWH}/g^{SM}_{WWH}$ as we scan over the $\xi-\beta_H$ parameters in the two 
regions?  The results of such a scan are shown in Fig.~\ref{fig1} for 
both regions I and II in units of $v/v_{SM}$ which we expect to be O(1); 
these results have been subjected to the additional 
requirement that the tachyon root, $x_T$, satisfy the bound $x_T <2$ so that the 
resulting Higgs mass lies below a TeV and, hence, is visible at LHC/ILC. Here we see 
that results in the two regions are quite different with much larger deviations 
from the SM seen in region I. It is important to note that not only are these 
deviations significant but they are restricted to a rather unique, narrow ranges 
in either region. With precision measurements at the ILC these predictions can be 
directly tested. The ratio $v/v_{SM}$ itself can be independently 
determined by combining the 
Higgs mass measurement with that of the 4-d quartic coupling via the 
relation $\lambda_{4d}=m_H^2/(2v^2)$. The first KK Higgs 
excitation may be light enough to be produced at a 1 TeV ILC so that 
it is important to examine these couplings as well which can be obtained in a 
manner similar to the above. The corresponding results can be seen in 
Fig.~\ref{fig2} for both regions where we see the previous pattern repeated. The 
reduced values of these couplings will lead to a substantial reduction in the 
production cross section for this state making it difficult to produce at LHC/ILC. 
\begin{figure*}[t]
\centerline{
\includegraphics[width=5.8cm,angle=-90]{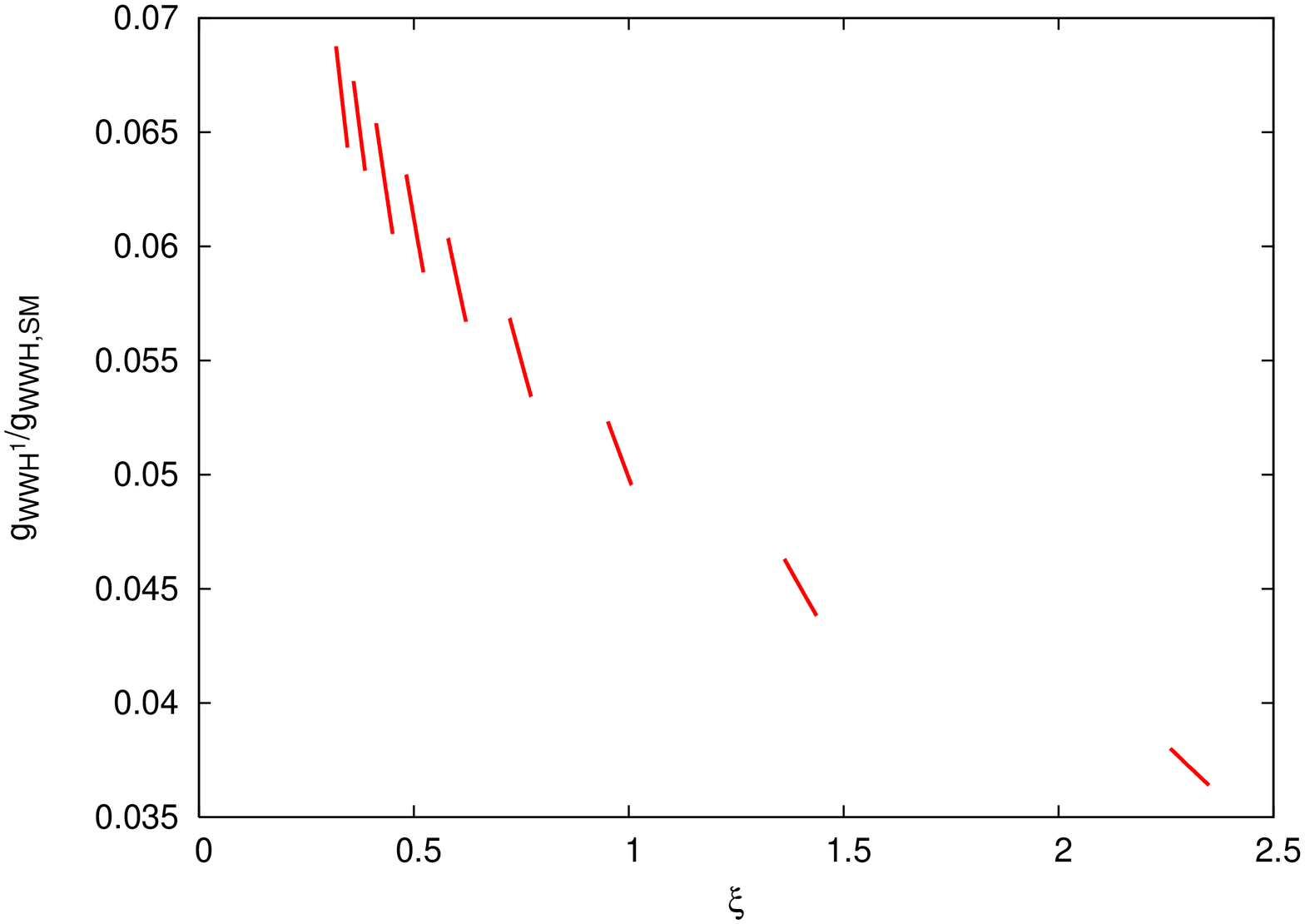}
\hspace*{5mm}
\includegraphics[width=5.8cm,angle=-90]{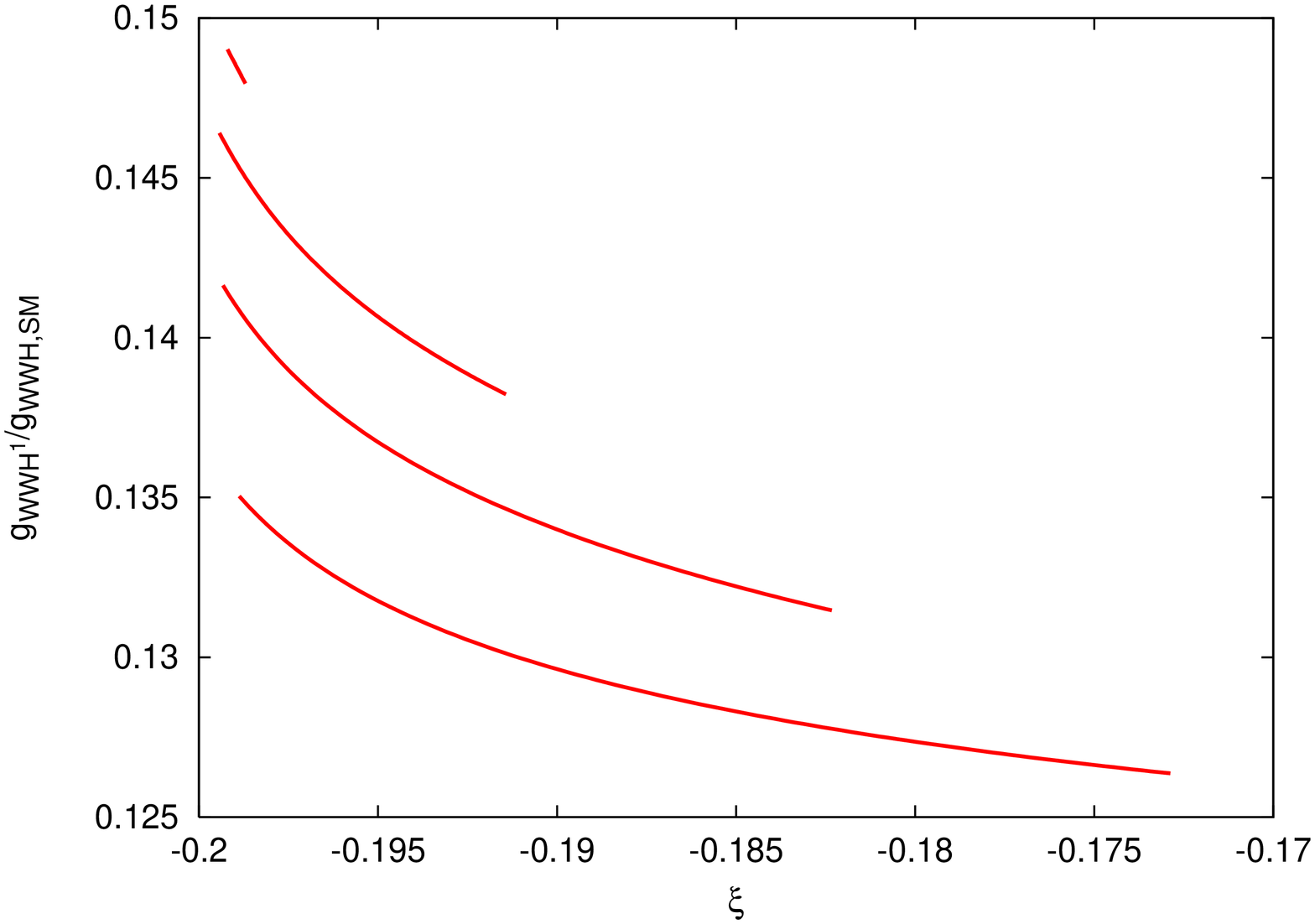}}
\vspace*{0.1cm}
\caption{Same as the previous figure but now for the first Higgs KK excitation.}
\label{fig2}
\end{figure*}

In the ``gravity-induced'' version of the URSM there are additional predictions which may 
be testable at the ILC. In this framework the parameter $\xi$ controls the size of 
Higgs-radion mixing which is now directly correlated with the $HWW$ coupling in the weak 
eigenstate basis. Measurements of radion properties will provide further tests of the 
URSM scenario.

\begin{acknowledgments}
The authors wish to thank the Aspen Center of Physics for their hospitality. 
Work supported by Department of Energy contracts DE-AC02-76SF0051   
and DE-FG02-95ER40896.
\end{acknowledgments}

\end{document}